# Supporting physics instructors to use a variety of evidence-based approaches to improve student learning: An example from quantum mechanics

Paul Justice[1], Emily Marshman[2], and Chandralekha Singh[3]

[1]*Department of Physics, University of Cincinnati, Cincinnati, OH 45221*

[2]*Department of Physics, Community College of Allegheny County, Pittsburgh, PA, 15212*

[3]*Department of Physics and Astronomy, University of Pittsburgh, Pittsburgh, PA, 15260*

**Abstract.** Physics instructors need support to successfully adopt and adapt evidence-based active engagement (EBAE) approaches because improving teaching and learning is a process and support is needed to ensure that they do not get disheartened if a particular EBAE approach does not produce the desired outcome. The instructors not only need support to refine their implementation of a specific EBAE approach to make them effective, but also to use a variety of EBAE methods to improve student learning. Here we illustrate how, with appropriate support, a quantum mechanics instructor did not give up when an EBAE approach involving implementation of a sequence of clicker questions on addition of angular momentum did not yield expected learning outcomes. The support ensured that the instructor remained optimistic and used another EBAE method that did not require him to spend more time in class on this topic. In particular, the instructor created an opportunity for students to productively struggle with the same problems (they had not performed well on after clicker questions) by giving them grade incentives to correct their mistakes outside of class. Student performance on one of the addition of angular momentum problems posed on the final exam suggests that students who corrected their mistakes benefited from the task and learned about addition of angular momentum better than those who did not correct their mistakes. Encouraging and supporting physics instructors can even be accomplished using an online community of physics educators. This type of support can go a long way in helping students learn physics because it is likely to increase their persistence in using various EBAE approaches as they refine their implementation to suit their students as well as their own instructional style.

## I. INTRODUCTION

One major cause for the low *sustained* usage of the evidence-based active engagement (EBAE) learning tools and pedagogies among physics instructors is that the early adaptations in their classes do not necessarily show the same large gains in student learning as those observed by their developers [1-4]. One reason for this failure to replicate in their own classes the positive outcomes of an innovation in early implementations is that the EBAE approaches must be refined to suit, e.g., the instructor's style and their students' prior knowledge and skills. If an EBAE pedagogy is adapted but does not produce the desired learning outcome early on, instructors may feel disappointed and quit using that innovative EBAE approach. Therefore, physics educators must be supported to remain persistent and recognize that improving teaching and learning is a process and successive refinements of EBAE approaches can lead to positive learning outcomes even when they do not appear to be successful in improving student learning in earlier implementations. The support provided to physics instructors by similar minded instructors can also be invaluable in helping them adapt several EBAE approaches for their students to meet the needs of their students as well as the constraints of their classroom to help their students master physics concepts.

Learning quantum mechanics (QM) is challenging even for advanced students partly due to the counter-intuitive nature of formalism, which is very different from classical mechanics [5-7]. Prior studies suggest that many upper-level undergraduate and graduate students struggle while learning QM [8, 9] and research-based approaches and learning tools can help students learn quantum concepts better [10-14]. Many researchers have developed conceptual surveys for evaluating student understanding of various quantum concepts [15-20] while others have investigated other aspects of learning QM, such as epistemological issues or student overconfidence [21, 22]. Also, to develop expertise, students must develop a functional understanding of different representations commonly used in QM such as the Dirac notation [23-34]. Moreover, the second quantum revolution and its potential in transformative quantum technologies entails focus on quantum information science and technology more broadly, e.g., see [35-51].

Some prior investigations have focused on helping students visualize quantum concepts [52-60], learn QM via games [61], while others have investigated instructor and teaching assistants' attitudes regarding teaching QM including their grading approaches [62, 63]. Other researchers have focused on student difficulties after traditional instruction including those related to quantum measurement [64-69], probability distributions for measuring physical observables, expectation values and their time dependence, student understanding of relative phase in the quantum states [70-74] as well as quantum experiments [75-80].

Other research studies have focused on cognitive issues, reasoning difficulties as well as transfer of learning specifically focusing on how they relate to the novel paradigm of QM [81-88]. Some researchers have focused on investigating student difficulties with other quantum concepts such as bound and scattering states, tunneling and conductivity [89-91] while other researchers have investigated student difficulties with finding a good basis for degenerate perturbation theory, fine structure corrections to hydrogen atom, Zeeman effect, identical particles and use of partial differential equations in QM [92-101] and developed research-based instructional materials and pedagogies to help students learn quantum concepts better.

This paper expands on a PERC proceedings [102] and provides an example from a quantum mechanics class to illustrate that with support from other physics educators versed in several EBAE approaches, a quantum physics instructor did not give up when an EBAE method involving implementation of a clicker [13, 103-113] question sequence (CQS) on addition of angular momentum did not yield expected learning outcomes on the posttest administered after the CQS. Instead, with support, the instructor viewed improving teaching and learning as a process, and with the same set of students, implemented another EBAE method in homework that did not require him to spend more time in class on this topic. In particular, the instructor provided grade incentives to students to correct their mistakes [14, 114-116], on the posttest before providing the correct solution, creating an out of class opportunity for students to productively struggle with the same problems on addition of angular momentum that they had not solved correctly. The instructor's use of this pedagogy, "incentives for learning from mistakes" [14, 114-116], appears to be valuable in that students' performance on one of the addition of angular momentum problems posed on the final exam showed that students who corrected their mistakes benefited from the task and performed better than those who did not correct their mistakes. We conclude this paper by emphasizing that having a local or online community of physics educators supporting and aiding each other in use of various EBAE approaches as they refine their implementation can be invaluable for helping their students learn physics.

**Background on "clicker" pedagogy:** This study is in the context of a quantum mechanics course in which instructors first used an EBAE pedagogy involving sequences of clicker questions. Clicker questions are conceptual multiple-choice questions typically administered in the classroom to engage students in the learning process and obtain feedback about their learning via a live feedback system called clickers [13, 103-113]. Integration of peer interaction with lectures via clicker questions has been popularized in the physics community by Mazur [103]. In a typical implementation, the instructor poses conceptual, multiple-choice clicker questions to students, which are interspersed throughout the lecture with the instructor receiving their responses. Students first answer each clicker question individually, which requires them to take a stand regarding their thoughts about the concept(s) involved. Students then discuss their answers to the questions with their peers and learn by articulating their thought processes and assimilating their thoughts with those of the peers. Then after the peer discussion, they answer the question again using clickers, followed by a general class discussion about those concepts in which both students and the instructor participate. The feedback that the instructor obtains is also valuable because it provides an estimate of the prevalence of student common difficulties and the fraction of the class that has understood the concepts and can apply them in the contexts in which the clicker questions are posed. The use of CQS keeps students alert during lectures and helps them monitor their learning. Clicker questions can be used in the classroom in different situations, e.g., they can be interspersed within lectures to evaluate student learning in each segment of a class focusing on a concept, at the end of a class, or to review materials from previous classes at the beginning of a class.

**Background on "incentives for learning from mistakes" (ILM) pedagogy:** Another EBAE pedagogy that a physics instructor used in this study involves giving grade incentives to students to struggle and learn from their own mistakes, when the CQS did not yield the desired outcome on the posttest after the CQS implementation. To appreciate the ILM pedagogy, we must recognize that a characteristic of physics experts is that they use problem-solving as an opportunity for learning. In particular, experts automatically reflect upon their mistakes in their problem solution to repair, extend and organize their knowledge structure. However, for many students, problem-solving can be a missed learning opportunity [14, 114-116]. Without guidance, students often do not necessarily reflect upon the problem-solving process after solving problems to learn from them nor do they try to learn from their mistakes after the graded problems are returned to them [14, 114-116] The ILM pedagogy is based on the tenet that instructors can explicitly incentivize students to learn from their mistakes by rewarding them for correcting their mistakes [14, 114-116]. This type of activity over time may also help students learn to make use of problem solving as a learning opportunity even if no grade incentive is provided.

Here we discuss how with encouragement and support, an instructor implemented the ILM pedagogy in a junior/senior level quantum mechanics course when a CQS did not yield the learning gains expected by the instructor. The ILM pedagogy proposes giving grade incentives to students for learning from their mistakes, e.g., by explicitly rewarding them for correcting their mistakes before giving them the correct solutions, because productive struggle while diagnosing one's mistakes and learning from them can be an excellent opportunity both for learning physics content and developing problem-solving skills. Students may gain a new perspective on their mistakes by asking themselves reflective questions while solving the problems correctly making use of the resources, e.g., their class notes and textbook available to them. In a prior study spanning four years in undergraduate quantum mechanics course, it was found that the incentive to correct the mistakes on the midterm exam had the greatest positive impact on the final exam performance of students who did poorly on the midterm exam compared to others [14].

## II. ADDITION OF ANGULAR MOMENTUM CQS

Before we discuss how the CQS on addition of angular momentum in quantum mechanics (QM) was implemented by the instructors, we summarize its development and validation process including its learning goals. This CQS was developed for students in upper-level undergraduate QM courses by taking advantage of the learning goals and inquiry-based guided learning sequences in a tutorial on this topic [117] as well as by refining, fine-tuning, and adding to the existing clicker questions from our group.

One learning goal of the CQS is that students should be able to identify the dimensionality of the product space of the spin of two particles. For example, if a system consists of two spin-1 particles with individual three-dimensional Hilbert spaces, the product space of the two-spin system is the product of those dimensions, 3×3=9 (not the sum of dimensions, 3+3=6). Another learning goal of the CQS is that students can choose a suitable representation, such as the "uncoupled" or "coupled" representation and construct a complete set of basis states for the product space in that representation. We note that the concepts related to the addition of orbital and spin angular momenta are analogous so here we will only focus on spin. In standard notation, the basis states in the uncoupled representation are eigenstates of $\hat{S}_1^2$, $\hat{S}_{z1}$, $\hat{S}_2^2$ and $\hat{S}_{z2}$ and can be written as $|s_1, m_{s1}\rangle \otimes |s_2, m_{s2}\rangle$. Here each particle's individual spin and z-component of spin quantum numbers are $s_1, s_2$ and $m_{s1}, m_{s2}$, respectively. On the other hand, in the coupled representation, the basis states, $|s, m_s\rangle$, are eigenstates of $\hat{S}_1^2$, $\hat{S}_2^2$, $\hat{S}^2$, and $\hat{S}_z$ where $\vec{\hat{S}} = \vec{\hat{S}}_1 + \vec{\hat{S}}_2$ and the total spin quantum number, s, and the z-component of the spin quantum number, $m_s$, are for the entire system. Students should be able to use the addition of angular momentum to determine that the total spin quantum number of the system s can range from $s_1 + s_2$ down to $|s_1 - s_2|$, with integer steps in between, where $s_1$ and $s_2$ are the individual spin quantum numbers for the particles. The z-component of the spin of the composite system is $m_s = m_{s1} + m_{s2}$. Another learning goal of the CQS is that students be able to calculate matrix elements of various operators corresponding to observables (e.g., a Hamiltonian in the product space) in different representations.

Based upon the learning goals, questions on the addition of angular momentum CQS were developed or adapted from prior validated clicker questions and sequenced to balance difficulties, avoid change of both concept and context between adjacent questions as appropriate to avoid cognitive overload [118], and include a mix of abstract and concrete questions to help students develop a good grasp of the concepts. The validation was an iterative process. After the initial development of the CQS using the learning goals and inquiry-based guided sequences in the tutorial and existing individually validated CQSs, we iterated the CQS with three physics faculty members who provided valuable feedback on fine-tuning and refining both the CQS as a whole and some new questions that were developed and adapted with existing ones to ensure that the questions were unambiguously worded and build on each other based upon the learning goals. We then conducted individual think-aloud interviews with advanced students who had learned these concepts via traditional lecture-based instruction in relevant concepts to ensure that they interpreted the CQS questions as intended and the sequencing of the questions provided the appropriate scaffolding support to students. This version of the CQS had 11 questions, which can be grouped into three sections (to be discussed below) and can be integrated with lectures in which these relevant concepts are covered in a variety of ways based upon the instructor's preferences. Later two more questions were added to the CQS.

The addition of angular momentum CQS has three sections that can be used separately or together depending, e.g., upon whether these are integrated with lectures like Mazur's approach, used at the end of each class or used to review concepts after students have learned via lectures everything related to addition of angular momentum that the instructor wanted to teach. The first section of the CQS, CQ1-CQ3, focuses on the uncoupled representation with basis states $|s_1 m_{s1}\rangle \otimes |s_2, m_{s2}\rangle$. The first question focuses on student understanding of the notation for the basis states in this representation along with the dimensionality of the product space. Following this question, CQ2 and CQ3 build on this understanding, asking students to identify the operators for which the basis states in the uncoupled representation are eigenstates and about some diagonal and off-diagonal matrix elements of various operators and whether they are zero or non-zero (i.e., determining whether operators are diagonal in the uncoupled representation). This section of the CQS concludes with a class discussion in which the instructor may review characteristics of this representation, as well as address any common difficulties exhibited by students.

The second section of this CQS, CQ4-CQ6, deals with the coupled representation with basis states $|s, m_s\rangle$ (where $s_1$ and $s_2$ are suppressed). The structure and concepts in these questions shown below are analogous to the structure of the first section, allowing students to compare and contrast these two representations. Below, correct answers are in bold.

*(CQ4)* Choose all of the following statements about the product space for a system of **two spin-1/2 particles** in the **coupled representation** that are correct:

  I. The dimensionality of the product space is the product of the dimensions of each particle's subspace, which is 2x2=4.
 II. $|s, m_s\rangle$ is an appropriate form for the basis states, where s ranges from $|s_1-s_2|$ to $s_1+s_2$ by integer steps, and $m_s=m_{s1}+m_{s2}$, ranging from –s to s in integer steps for each s.
III. $|1,1\rangle$, $|1,0\rangle$, $|1,-1\rangle$, and $|0,0\rangle$ are the elements of a complete set of basis states.

   a) I only  b) I and II only  c) I and III only
   d) II and III only  **e) All of the above**

---

*(CQ5)* Choose all of the following statements about the product space for a system of **two spin-1/2 particles** in the **coupled representation** that are correct:
  I. Basis state $|1,-1\rangle$ is an eigenstate of $\hat{S}^2$ such that $\hat{S}^2|1,-1\rangle = 2\hbar^2|1,-1\rangle$.
 II. Basis state $|1,-1\rangle$ is an eigenstate of both $\hat{S}_1^2$ and $\hat{S}_2^2$ such that $\hat{S}_1^2|1,-1\rangle = 2\hbar^2|1,-1\rangle$ and $\hat{S}_2^2|1,-1\rangle = 2\hbar^2|1,-1\rangle$.
III. Basis state $|1,-1\rangle$ is an eigenstate of $\hat{S}_{z1}$, $\hat{S}_{z2}$, and $\hat{S}_z$.

   a) **I only**  b) I and II only  c) I and III only
   d) II and III only  e) All of the above

---

*(CQ6)* Consider the product space of a system of **two spin-1/2 particles**. Choose all of the following that are correct regarding the scalar products. (Recall that these scalar products are the matrix elements of the $\hat{S}_{1z} + \hat{S}_{2z}$ operator in this basis).

  I. $\langle 1,1|(\hat{S}_{z1} + \hat{S}_{z2})|1,0\rangle = \langle 1,1|\hat{S}_z|1,0\rangle = 0$
 II. $\langle 1,-1|(\hat{S}_{z1} + \hat{S}_{z2})|1,-1\rangle = \langle 1,-1|\hat{S}_z|1,-1\rangle = -\hbar$
III. $(\hat{S}_{z1} + \hat{S}_{z2})$ is diagonal in the coupled representation.
 IV. $(\hat{S}_{z1} + \hat{S}_{z2})$ is diagonal in the uncoupled representation.

   a) II and III only  b) I, II, and III only  c) I and IV only
   d) I, II, and IV only  **e) All of the above**.

---

As noted, the first two sections of the addition of angular momentum CQS (except CQ6) deal with only one representation at a time, and only with a system of two spin-1/2 particles. This choice is deliberate by design to avoid cognitive overload and allow students to revisit these representations in a familiar context since typical instruction on these concepts tends to emphasize a system of two spin-1/2 particles first.

The third section of the CQS extends these concepts to higher dimensional product spaces for both coupled and uncoupled representations. For example, CQ7 deals with the dimensionalities of the product space for systems of two spins that are not both spin-1/2. Then, CQ8 and CQ9 ask students to identify basis states in the coupled and uncoupled representations for these less familiar two-spin systems. CQ10 focuses on a system of spin-1/2 and spin-1 particles in the coupled representation and CQ11 focuses on convenient representation for certain operators. Also, CQ12 and CQ13 ask students to identify the basis in the product space in which given Hamiltonians are diagonal. These Hamiltonians are comprised of operators addressed previously in the first questions of the CQS.

**In-Class Implementation of CQS by Instructor A**: The CQS was implemented with peer discussion [103] in an upper-level undergraduate QM class at a large research university after traditional lecture-based instruction in relevant concepts on the addition of angular momentum in which students learned about the coupled and uncoupled representations not only for a system of two spin-1/2 particles but also for systems for which the product spaces involve higher dimensions. Prior to the implementation of the CQS in class, students took a pretest after traditional lecture-based instruction in relevant concepts in each class, which was developed and validated by Zhu et al. [117] to measure comprehension of the concepts of addition of angular momentum. The first six questions in the CQS were implemented together right after the pretest. The last five questions in the third section of the addition of angular momentum CQS were implemented at the beginning of the next class to review concepts covered earlier in the lectures on product spaces involving higher dimensions. The posttest was administered during the following week to measure the impact of the CQS.

On the pretest, students were given a system of two spin-1/2 particles and a spin-spin interaction Hamiltonian, $\hat{H}_1 = \left(4E_0/\hbar^2\right)\hat{S}_1 \cdot \hat{S}_2 = \left(2E_0/\hbar^2\right)(\hat{S}^2 - \hat{S}_1^2 - \hat{S}_2^2)$, and a magnetic field-spin interaction Hamiltonian, $\hat{H}_2 = -\mu B(\hat{S}_{z1} + \hat{S}_{z2})$, and asked to answer these questions:

(a) *Write down a complete set of basis states for the product space of a system of two spin-1/2 particles. Explain the labels you are using to identify your basis states.*

(b) *Evaluate one diagonal and one off-diagonal matrix element of the Hamiltonian $\hat{H}_1$ (of your choosing) in the basis you have chosen. Label the matrix elements so that it is clear which matrix elements they are.*

(c) *Evaluate one diagonal and one off-diagonal matrix element of the Hamiltonian $\hat{H}_2$ (of your choosing) in the basis you have chosen. Label the matrix elements so that it is clear which matrix elements they are.*

(d) *Are both Hamiltonians $\hat{H}_1$ and $\hat{H}_2$ diagonal matrices in the basis you chose?*

The posttest that students were administered following the implementation of the CQS was analogous to the pretest and asked the same questions as the pretest but for a system of one spin-1/2 particle and one spin-1 particle. These pre/posttests are very similar to those administered by Zhu et al. to measure student learning after traditional instruction and after engaging with the addition of angular momentum tutorial [117]. However, due to time constraints in the classroom, questions (b) and (c), which had previously asked students to construct the entire matrix representation of the Hamiltonians, were reduced as stated earlier to evaluation of only one diagonal and off-diagonal matrix element [117]. To compare the performance of CQS and tutorial groups on pre/posttests so that the relative improvements can be determined, the same rubric was used for pre-/posttests given to the CQS students as the tutorial students in Ref. [117] (who were also advanced undergraduate students in QM). Questions (a), (b), and (c) were each worth 3 points, and students were awarded partial credit if only some basis states in (a) or some matrix elements in (b) or (c) were correct. Question (d) was worth 1 point (correct answer "yes or no").

### III. IN-CLASS IMPLEMENTATION RESULTS FOR CQS BY INSTRUCTOR A IN CLASS A

Tables 1 and 2 compare pre/posttest performances of upper-level QM students from the same university in two different years after traditional lecture-based instruction (pretest) and on posttest after students had engaged with the CQS (Table 1) or tutorial (Table 2) on the addition of angular momentum. The normalized gain (or gain) is calculated as $g = (post\% - pre\%)/(100\% - pre\%)$ [119] and presented in both Tables 1 and 2 but effect size is calculated only in Table 1 (not available for Table 2 data in Ref. [117]). Effect size was calculated as Cohen's $d = (\mu_{post} - \mu_{pre})/\sigma_{pooled}$ where $\mu_i$ is the mean of group $i$ and $\sigma_{pooled}$ is the pooled standard deviation [120].

**Table 1.** Comparison of mean pre/posttest scores on each question, normalized gains and effect sizes for upper-level undergraduate QM students in class A who engaged with the CQS on addition of angular momentum concepts (N=16).

| Question | Pretest Mean | Posttest Mean | Normalized Gain (g) | Effect Size (d) |
|---|---|---|---|---|
| (a) | 59% | 95% | 0.88 | 0.30 |
| (b) | 24% | 48% | 0.31 | 0.22 |
| (c) | 17% | 71% | 0.66 | 0.44 |
| (d) | 14% | 43% | 0.33 | 0.67 |
| Total | 31% | 69% | 0.54 | 0.35 |

**Table 2.** Comparison of mean pre/posttest scores on each question and normalized gains from Ref. **[117]** (effect sizes not available) for upper-level undergraduate QM students who engaged with the tutorial on addition of angular momentum concepts (N=26).

| Question | Pretest Mean | Posttest Mean | Normalized Gain (g) |
|---|---|---|---|
| (a) | 77% | 85% | 0.35 |
| (b) | 8% | 54% | 0.50 |
| (c) | 8% | 73% | 0.71 |
| (d) | 31% | 85% | 0.78 |
| Total | 34% | 72% | 0.58 |

Although the number of students in each class is small and the pretest scores in Tables 1 and 2 are often different, they are low in both tables (except for question (a) in Table 2). However, the comparison of the posttest scores of the CQS group and the tutorial group in Tables 1 and 2 suggests that the CQS is effective in helping students learn to construct a complete set of basis states (question (a)) and calculate matrix elements for the magnetic field-spin interaction Hamiltonian (question (c)), garnering similar posttest scores to those of students who engaged with the tutorial. However, Table 1 also shows that students did not perform well on questions (b) and (d) even after engaging with the CQS. Review of student responses suggests that a major reason for the poor performance on both of these questions, even after the CQS, is due to the fact that a majority of students chose the basis to be the uncoupled representation (since it is the simpler representation for constructing the basis states) and then had difficulty with the matrix elements of the spin-spin interaction Hamiltonian in questions (b) and (d) since it is only diagonal in the coupled representation. In particular, in question (a), many students correctly constructed a complete set of basis states but chose the uncoupled representation.

We note that while the magnetic field-spin interaction Hamiltonian in question (c) is diagonal in both coupled and uncoupled representations, calculating the matrix elements of the spin-spin interaction Hamiltonian in question (b) in the uncoupled representation is challenging since that operator is not diagonal in this basis. Along with a reasonable posttest score for question (a), the CQS group students' poor posttest score on questions (b) and (d) in Table 1 may be due to the fact that while students learned to construct a complete set of basis states, many were not versed in calculating the matrix elements of an operator in a representation in which it is not diagonal as in question (b) (many students assumed that the spin-spin Hamiltonian in question (b) is also diagonal in the uncoupled representation, which it is not).

In fact, for question (d), even after the CQS, many students claimed that both Hamiltonians are diagonal in the uncoupled representation they had chosen. Since students were only asked to calculate a single off-diagonal matrix element in question (b), some students who correctly calculated an off-diagonal matrix element in question (b) that was zero concluded that the entire $\hat{H}_1$ matrix is diagonal in the uncoupled representation which it is not. On the other hand, a comparison of student performances on posttest in Tables 1 and 2 for questions (b) and (d) suggests that most students who engaged with the tutorial answered question (d) correctly but struggled to calculate matrix elements on the posttest in question (b).

## IV. COMBINING OF TWO EBAE PEDAGOGIES BY INSTRUCTOR B

As noted, physics instructors must be provided support for staying optimistic and refining an EBAE approach or incorporating more than one approach as needed to help students learn. While EBAE strategies are likely to provide promising results in a classroom after some adaptations, instructors must recognize that improvement in teaching and learning is a process that may not yield the desired outcome particularly in the first few implementations. Thus, the implementation of the EBAE methods needs to be refined to suit instructors' teaching style and their students' prior knowledge and skills. For this reason, it is important for instructors to be supported in keeping and employing several EBAE instructional tools from their toolbox.

Following the implementation of the CQS in class A, which yielded reasonably good performance on two posttest questions but not on the other two posttest questions after students engaged with the CQS, two more clicker questions, CQ10 and CQ11, were added based upon the difficulties found after implementation in class A. This slightly amended CQS was then implemented in class B by instructor B, who was a different instructor than that for class A. The implementation of the CQS in class B followed the same procedure discussed in the preceding section for class A.

Table 3 shows class B's performance on all parts of both the pretest and posttest. Table 3 shows that students' average posttest performance was poor except on question (a). Students in this class performed significantly worse than class A (in Table 1) even on question (c). Although the normalized gains and effect sizes on all questions are reasonable (see Table 3), the instructor of course B was concerned about the learning as measured by the posttest scores and the fact that most students had not mastered the concepts. Although how the CQS was implemented in class B could have played an important role in why the students did not benefit significantly from it, one likely reason for not benefiting from the CQS is that students did not have sufficient initial knowledge (as evidenced by the pretest scores) before they engaged with the CQS. For example, when roughly half of the students know the correct answers to the clicker questions, the peer discussions during the implementation of the clicker questions is generally effective [103]. One possible reason for the low prior preparation as evidenced by the pretest scores may be that the instructor of class B did not spend sufficient time before the CQS on discussing the relevant underlying concepts (e.g., on questions (b) and (d), students in class B performed very poorly on the pretest as shown in Table 3 and for those questions their posttest scores are also less than 40%).

**Table 3.** Comparison of mean pre/posttest scores on each question, normalized gains and effect sizes for upper-level undergraduate QM students in class B who engaged with the CQS on addition of angular momentum concepts (N=19).

| Part | Pretest Mean | Posttest Mean | Normalized Gain (g) | Effect Size (d) |
|---|---|---|---|---|
| (a) | 46% | 86% | 0.74 | 0.26 |
| (b) | 5% | 39% | 0.35 | 0.30 |
| (c) | 12% | 53% | 0.46 | 0.30 |
| (d) | 3% | 34% | 0.32 | 0.66 |
| Total | 19% | 57% | 0.46 | 0.30 |

Following these posttest results on addition of angular momentum, the instructor was encouraged and provided support to implement another active learning pedagogy. In particular, the instructor used the ILM pedagogy and returned the posttests to students with grades and incorrect parts marked but without explanations and asked them to correct their mistakes as homework in return for up to half of the quiz points they had lost. Unlike the earlier implementation of the ILM pedagogy in quantum mechanics in which students were asked to correct their mistakes on midterm exams with similar incentives to earn 50% of the missed points (in which case all students corrected their mistakes), not all students took advantage of the opportunity because the posttest was a low stakes quiz worth less than 1% of a student's final grade in the course. Table 4 shows the results after 12 of the 19 students made corrections to their posttests. With implementations of both the CQS and learning from mistakes pedagogies, students in class B who corrected their mistakes demonstrated better performance (see Table 4). Moreover, we note that instructor B gave question (c) as part of the midterm exam. After corrections, students who corrected their mistakes on the posttest obtained an average score of 85% on this problem. Meanwhile, students who did not correct their posttest obtained an average score of 71% (note that those who corrected their mistakes initially had a slightly lower score on this question than those who did not correct their mistakes). While students who did not correct their mistakes also performed better on the midterm exam (71%) since they also had access to the correct solution and had further opportunity to learn concepts, those who corrected their mistakes performed significantly better than them (85%).

Table 4. Comparison of mean score on each question before and after student corrections for upper-level undergraduate QM students in class B who engaged with the CQS on addition of angular momentum concepts and engaged with the ILM pedagogy to learn from their mistakes. Columns showing only students who made corrections (N=12) are shown alongside the class average (N=19). Last column shows how students performed when they were asked to correct their mistakes.

| Part | Initial Posttest | | | After Corrections |
|---|---|---|---|---|
| | Correctors (N=12) | Non-Correctors (N=7) | All Participants (N=19) | Correctors (N=12) |
| (a) | 94% | 71% | 86% | 100% |
| (b) | 47% | 24% | 39% | 94% |
| (c) | 50% | 57% | 53% | 86% |
| (d) | 25% | 50% | 34% | 67% |
| Total | 60% | 51% | 57% | 91% |

## V. DISCUSSION AND SUMMARY

Physics instructors who adapt EBAE approaches in their classes often give up if they do not yield desirable learning outcomes in early implementations. Therefore, physics instructors should be supported to view improvement in their teaching and learning as a process and adapt different instructional approaches with the same set of students. Encouraging and supporting physics instructors to keep several EBAE learning tools and pedagogies in their toolbox is critical to improve student learning. With appropriate support, instructors can combine various EBAE approaches if the results from the implementation of one of these EBAE approaches do not yield the desired outcome. Ensuring that the details of the implementation of the chosen approach are well-matched with their style and their students' prior knowledge and skills may take time. In particular, different physics instructors may have students with different prior preparations, their teaching styles may be different, and they may also have varying degrees of familiarity with the EBAE tools they are using. Thus, some tools may prove to be less effective than anticipated at least in earlier implementations and an iterative approach with successive refinements may be required to get them closer to their instructional goal. Moreover, when an approach does not yield the desired learning outcome, physics educators should be provided support to improvise using additional EBAE methods commensurate with the constraints of their physics courses to improve learning of students, who may not have benefited from one approach.

We presented an example of two quantum physics instructors, who first used a CQS on addition of angular momentum. The CQS was implemented by the two instructors, A and B, in the same QM course in consecutive years. After the in-class implementation of the CQS in Class A, it was found that the CQS was effective in helping students construct a complete set of basis states in a product space and in calculating matrix elements for an operator that is diagonal in that basis, but not on all concepts. The CQS was slightly modified based upon this feedback and was implemented the following year at the same institution by instructor B in QM Class B. Student performance on many of the questions was worse than those of instructor A's students suggesting insufficient mastery of the concepts even after the CQS implementation. This is not an uncommon occurrence for instructors adopting new instructional tools, as the instructional tool must be adapted to the instructor's style as well as students' prior preparation. The corresponding result shows that in this case, both before and after the CQS implementation, class B's conceptual understanding was lagging relative to Class A, so with encouragement and support, instructor B adapted to his class's needs using the ILM pedagogy that did not require devoting more class time on these concepts. In particular, with encouragement from others versed in pedagogy, he used the ILM pedagogy and gave students grade incentives to correct their mistakes on the posttest. Students who took advantage of this opportunity and made posttest corrections not only showed gains on the corrected posttest, but they also performed better on a final exam question that focused on the same concepts. We note that even students who did not make corrections to their posttest had the opportunity to take advantage of learning from the solutions to the posttest provided for the class after students had the opportunity to correct their mistakes. However, their average score on the final exam question was lower than that of the group that took advantage of the ILM pedagogy and corrected their mistakes on the posttest. This finding is consistent with the previous study involving the ILM pedagogy in which advanced undergraduate physics students in a similar QM course performed better on related tasks after being given incentives to correct their mistakes [14].

In summary, physics educators should be provided encouragement and support for using various EBAE pedagogies as appropriate to suit the instructional situation at a given time and view teaching as a process because this can be invaluable in improving physics teaching and learning. In the study discussed, discussions with other physics instructors versed in EBAE approaches led instructor B in QM Class B to incorporate the ILM pedagogy, when a CQS did not show as much benefit as anticipated. Furthermore, since local support may not be available to physics educators at all institutions, it would be valuable to have online communities of physics educators across different colleges and universities. This community of physics educators can encourage and support each other so that instructors whose EBAE pedagogical implementation was not effective recognize that improving teaching is a process and do not get disheartened as they become comfortable using multiple EBAE pedagogies in their physics courses. While this mutual support community of physics educators can communicate via an asynchronous platform such as Discord, a monthly synchronous meeting of physics instructors with similar interests on a platform like Zoom, e.g., those teaching QM can be valuable in helping our students develop a functional understanding of physics.

## ACKNOWLEDGEMENT

We thank the National Science Foundation for award PHY-2309260. We thank all students and faculty members who helped with this study and to R. P. Devaty for helpful feedback.

## APPENDIX: CQS ON THE ADDITION OF ANGULAR MOMENTUM

*(CQ1)* *Choose all of the following statements about the product space for a system of **two spin-1/2 particles** in the **uncoupled representation** that are correct:*

I. The dimensionality of the product space is the sum of dimensions from each particle's subspace. That is 2+2=4.
II. $|s_1 m_{s1}\rangle \otimes |s_2, m_{s2}\rangle$ is an appropriate form for the basis vectors.
III. $\left|\frac{1}{2}, \frac{1}{2}\right\rangle \otimes \left|\frac{1}{2}, \frac{1}{2}\right\rangle$, $\left|\frac{1}{2}, \frac{-1}{2}\right\rangle \otimes \left|\frac{1}{2}, \frac{-1}{2}\right\rangle$, and $\left|\frac{1}{2}, \frac{1}{2}\right\rangle \otimes \left|\frac{1}{2}, \frac{-1}{2}\right\rangle$ are a complete set of basis vectors.

b) I only  **b) II only**  c) III only
d) II and III only  e) None of the above

---

*(CQ2)* *Choose all of the following statements about the product space for a system of **two spin-1/2 particles** in the **uncoupled representation** that are correct:*

I. Basis vector $\left|\frac{1}{2}, \frac{-1}{2}\right\rangle \otimes \left|\frac{1}{2}, \frac{1}{2}\right\rangle$ is an eigenstate of $\hat{S}_{1z}$ such that:
$$\hat{S}_{1z} \left|\frac{1}{2}, \frac{-1}{2}\right\rangle \otimes \left|\frac{1}{2}, \frac{1}{2}\right\rangle = \left(\hat{S}_{1z} \otimes \mathbb{I}\right)\left(\left|\frac{1}{2}, \frac{-1}{2}\right\rangle \otimes \left|\frac{1}{2}, \frac{1}{2}\right\rangle\right) = \left(\hat{S}_{1z} \left|\frac{1}{2}, \frac{-1}{2}\right\rangle\right) \otimes \left(\mathbb{I} \left|\frac{1}{2}, \frac{1}{2}\right\rangle\right) = \frac{-\hbar}{2} \left|\frac{1}{2}, \frac{-1}{2}\right\rangle \otimes \left|\frac{1}{2}, \frac{1}{2}\right\rangle.$$

II. Basis vector $\left|\frac{1}{2}, \frac{-1}{2}\right\rangle \otimes \left|\frac{1}{2}, \frac{1}{2}\right\rangle$ is an eigenstate of $\hat{S}_{2z}$ such that:

$$\hat{S}_{2z}\left|\frac{1}{2},\frac{-1}{2}\right\rangle \otimes \left|\frac{1}{2},\frac{1}{2}\right\rangle = (\mathbb{I} \otimes \hat{S}_{2z})\left(\left|\frac{1}{2},\frac{-1}{2}\right\rangle \otimes \left|\frac{1}{2},\frac{1}{2}\right\rangle\right) = \left(\mathbb{I}\left|\frac{1}{2},\frac{-1}{2}\right\rangle\right) \otimes \left(\hat{S}_{2z}\left|\frac{1}{2},\frac{1}{2}\right\rangle\right) = \frac{\hbar}{2}\left(\left|\frac{1}{2},\frac{-1}{2}\right\rangle \otimes \left|\frac{1}{2},\frac{1}{2}\right\rangle\right).$$

III. Basis vector $\left|\frac{1}{2},\frac{-1}{2}\right\rangle \otimes \left|\frac{1}{2},\frac{1}{2}\right\rangle$ is an eigenstate of $\hat{S}_1^2$ and $\hat{S}_2^2$.

a) I only  b) II only  c) I and III only
d) II and III only  e) **All of the above**

---

*(CQ3)* Consider the product space of a system of **two spin-1/2 particles** in the **uncoupled representation**. In this representation, it is most useful to write $\hat{S}_1 \cdot \hat{S}_2$ as $\frac{\hat{S}_{1-}\hat{S}_{2+} + \hat{S}_{1+}\hat{S}_{2-}}{2} + \hat{S}_{1z}\hat{S}_{2z}$. Choose all of the following that are correct concerning scalar products. (Recall that these scalar products are the matrix elements of the $\hat{S}_{1-}\hat{S}_{2+}$ matrix in this basis).

I. $\left\langle \frac{1}{2},\frac{-1}{2}\right| \otimes \left\langle\frac{1}{2},\frac{1}{2}\right| \hat{S}_{1-}\hat{S}_{2+} \left|\frac{1}{2},\frac{1}{2}\right\rangle \otimes \left|\frac{1}{2},\frac{-1}{2}\right\rangle = (\left\langle\frac{1}{2},\frac{-1}{2}\right|\hat{S}_{1-}\left|\frac{1}{2},\frac{1}{2}\right\rangle)(\left\langle\frac{1}{2},\frac{1}{2}\right|\hat{S}_{2+}\left|\frac{1}{2},\frac{-1}{2}\right\rangle) = 0$

II. $\hat{S}_1 \cdot \hat{S}_2$ will be diagonal in the uncoupled representation

III. $\left\langle \frac{1}{2},\frac{-1}{2}\right| \otimes \left\langle\frac{1}{2},\frac{1}{2}\right| \hat{S}_{1-}\hat{S}_{2+} \left|\frac{1}{2},\frac{1}{2}\right\rangle \otimes \left|\frac{1}{2},\frac{-1}{2}\right\rangle = (\left\langle\frac{1}{2},\frac{-1}{2}\right|\hat{S}_{1-}\left|\frac{1}{2},\frac{1}{2}\right\rangle)(\left\langle\frac{1}{2},\frac{1}{2}\right|\hat{S}_{2+}\left|\frac{1}{2},\frac{-1}{2}\right\rangle) = \hbar^2$

IV. Some off-diagonal elements of $\hat{S}_1 \cdot \hat{S}_2$ in the uncoupled representation are non-zero.

a) II and III only  **b) III and IV only**  c) II and III only
d) I and IV only  e) None of the above.

---

*(Class Discussion Notes)*

Basis vectors in the uncoupled representation are eigenstates of $\hat{S}_{1z}$, $\hat{S}_{2z}$, $\hat{S}_1^2$, and $\hat{S}_2^2$.
Basis vectors: $\left|\frac{1}{2},\frac{1}{2}\right\rangle \otimes \left|\frac{1}{2},\frac{1}{2}\right\rangle$, $\left|\frac{1}{2},\frac{1}{2}\right\rangle \otimes \left|\frac{1}{2},\frac{-1}{2}\right\rangle$, $\left|\frac{1}{2},\frac{-1}{2}\right\rangle \otimes \left|\frac{1}{2},\frac{1}{2}\right\rangle$, and $\left|\frac{1}{2},\frac{-1}{2}\right\rangle \otimes \left|\frac{1}{2},\frac{-1}{2}\right\rangle$

*CQ4-CQ6 provided in the text are implemented here*

---

*(Class Discussion Notes)*

Basis vectors in the coupled representation are eigenstates of $\hat{S}_z$, $\hat{S}_1^2$, $\hat{S}_2^2$, and $\hat{S}^2$.
Basis vectors: $|1,1\rangle, |1,0\rangle, |1,-1\rangle,$ and $|0,0\rangle$

Compare and contrast uncoupled and coupled representations.

*(CQ7)* Choose all of the following that are correct:

I. The product space for a system of **a spin-1 particle and a spin-3/2 particle** is 3+4=7 dimensional in both coupled and uncoupled representations.
II. The product space for a system of **a spin-1/2 particle and a spin-1 particle** is 2x3=6 dimensional in the coupled representation, but not in the uncoupled representation.
III. The product space for a system of **two spin-1 particles** is 3x3=9 dimensional in both coupled and uncoupled representations.

a) I only  b) II only  **c) III only**
d) II and III only  e) None of the above

---

*(CQ8)* Choose all of the following that are correct about the basis vectors in the **<u>uncoupled</u>** representation.

I. If $s_1 = s_2 = 1$, $|1,-1\rangle \otimes |1,1\rangle$, $|1,\frac{1}{2}\rangle \otimes |1,\frac{1}{2}\rangle$, and $|1,0\rangle \otimes |1,1\rangle$ are appropriate basis vectors.

II. If $s_1 = s_2 = \frac{3}{2}$, $|0,0\rangle \otimes |1,1\rangle$, $|1,1\rangle \otimes |0,0\rangle$, and $|1,0\rangle \otimes |1,0\rangle$ are appropriate basis vectors.

III. If $s_1 = 1$, $s_2 = \frac{3}{2}$, $|1,0\rangle \otimes |\frac{3}{2},\frac{1}{2}\rangle$, $|1,0\rangle \otimes |\frac{3}{2},\frac{-3}{2}\rangle$, and $|1,1\rangle \otimes |\frac{3}{2},\frac{3}{2}\rangle$ are appropriate basis vectors.

a) I only  b) II only  **c) III only**
d) II and III only  e) None of the above

---

*(CQ9)* Choose all of the following that are correct about the basis vectors in the **coupled** representation:

I. For the product space of a system of **two spin-$\frac{3}{2}$ particles**, the possible values for the total spin quantum number s are 0, 1, 2, and 3, with some basis vectors being $|3,-1\rangle$, $|0,-1\rangle$, and $|1,0\rangle$.

II. For the product space of a system **of two spin-1 particles**, the possible values for s are 0, 1, and 2, with some examples of basis vectors being $|1,-1\rangle$, $|2,1\rangle$, and $|0,0\rangle$.

III. For the product space of a system of **a spin-1 particle and a spin-3/2 particle**, the possible values for s are 0, 1/2, 3/2, and 5/2, with some basis vectors being $|\frac{5}{2},\frac{-3}{2}\rangle$, $|\frac{1}{2},\frac{-1}{2}\rangle$, and $|\frac{3}{2},\frac{1}{2}\rangle$.

a) I only  **b) II only**  c) III only
d) II and III only  e) None of the above

---

*(CQ10)* Choose all of the following that are correct about the product space of a system of **a spin-$\frac{1}{2}$ particle and a spin-1 particle** with the basis vectors in the **coupled** representation:

I. The possible values for the total spin quantum number s are $\frac{1}{2}$ and $\frac{3}{2}$, with $m_s = \frac{1}{2}, \frac{-1}{2}, \frac{3}{2}, \frac{-3}{2}$ for $s = \frac{3}{2}$ and $m_s = \frac{1}{2}, \frac{-1}{2}$ for $s = \frac{1}{2}$.

II. This is a 4-dimensional product space with basis vectors $|\frac{1}{2},\frac{1}{2}\rangle$, $|\frac{1}{2},-\frac{1}{2}\rangle$, $|\frac{3}{2},\frac{3}{2}\rangle$, and $|\frac{3}{2},-\frac{3}{2}\rangle$ because the possible values of $m_s = \frac{1}{2}, \frac{-1}{2}, \frac{3}{2}, \frac{-3}{2}$.

III. This is a 6-dimensional product space with basis vectors $|\frac{1}{2},\frac{1}{2}\rangle$, $|\frac{1}{2},-\frac{1}{2}\rangle$, $|\frac{3}{2},\frac{3}{2}\rangle$, $|\frac{3}{2},-\frac{3}{2}\rangle$, $|\frac{3}{2},\frac{1}{2}\rangle$, and $|\frac{3}{2},-\frac{1}{2}\rangle$ because the values of $m_s = \frac{1}{2}, \frac{-1}{2}, \frac{3}{2}, \frac{-3}{2}$ for $s = \frac{3}{2}$ and $m_s = \frac{1}{2}, \frac{-1}{2}$ for $s = \frac{1}{2}$.

a) I only  b) II only  c) I and II only
**d) I and III only**  e) All of the above.

---

*(Class Discussion Notes)*

What are the values of the total spin quantum number and the z component of the total spin quantum number, s and $m_s$, respectively, for the product space of a system of **a spin-$\frac{1}{2}$ particle and a spin-1 particle**, e.g., for writing the basis states in the **coupled** representation?

What is the dimensionality of this product space?

---

*(CQ11)* When working with a given operator(s), it is useful to consider which representation is more convenient to work in. Choose all of the following statements that are correct about choosing a convenient basis to work in:

I. Basis vectors in the uncoupled representation are eigenstates of $\hat{S}_{1z}$, and $\hat{S}_{2z}$, making a convenient basis for operators that commute with $\hat{S}_{1z}$, and $\hat{S}_{2z}$.

II. Basis vectors in the coupled representation are eigenstates of $\hat{S}_z$ and $\hat{S}^2$, making a convenient basis for operators that commute with $\hat{S}_z$ and $\hat{S}^2$.

III. Basis vectors in both couple and uncoupled representations are eigenstates of $\hat{S}_1^2$ and $\hat{S}_2^2$, making either convenient basis for operators that commute with $\hat{S}_1^2$ and $\hat{S}_2^2$.

a) I only  b) II only  c) I and II only
d) I and III only  **e) All of the above**

---

**(CQ12)** Consider the Hamiltonian $\hat{H} = C\hat{S}_1 \cdot \hat{S}_2$. Choose all of the following statements that are correct about this operator acting on the basis vectors for the product space of a system of **two identical particles with non-zero spin:**

I. The Hamiltonian matrix is diagonal in the coupled basis because the basis vectors in the coupled representation are eigenstates of the operators $\hat{S}^2$, $\hat{S}_1^2$, and $\hat{S}_2^2$
II. The Hamiltonian matrix is diagonal in the uncoupled basis because the basis vectors in the uncoupled representation are eigenstates of the operators $\hat{S}_{1+}, \hat{S}_{1-}, \hat{S}_{2+}, \hat{S}_{2-}, \hat{S}_{1z}$, and $\hat{S}_{2z}$
III. For a system of **two identical spin-1/2 particles**, this Hamiltonian matrix is 4-by-4 in either the coupled representation or uncoupled representation.

a) I only  b) II only  c) I and II only
**d) I and III only**  e) All of the above.

---

*(Class Discussion Notes)*
For the Hamiltonian $\hat{H} = C\hat{S}_1 \cdot \hat{S}_2$, which representation is <u>convenient</u>? Uncoupled, coupled, both, or neither? Why?

For the Hamiltonian $\hat{H} = C\hat{S}_1 \cdot \hat{S}_2$, the Hamiltonian is diagonal in the **coupled representation**, but not the uncoupled representation. The **coupled representation is** <u>convenient</u>.

---

**(CQ13)** Consider the Hamiltonian $\hat{H} = C(\hat{S}_{1z} + \hat{S}_{2z})$. Choose all of the following statements that are correct about this operator acting on basis vectors for the product space of a system of **two identical particles with non-zero spin:**

I. The Hamiltonian matrix is diagonal in the coupled basis because the basis vectors in the coupled representation are eigenstates of the operator $\hat{S}_z = (\hat{S}_{1z} + \hat{S}_{2z})$
II. The Hamiltonian matrix is diagonal in the uncoupled basis because the basis vectors in the uncoupled representation are eigenstates of the operators $\hat{S}_{1z}$ and $\hat{S}_{2z}$
III. For a system of **two identical spin-1/2 particles**, this Hamiltonian matrix is 4-by-4 whether the coupled representation or uncoupled representation is chosen.

a) I only  b) II only  c) I and II only
d) I and III only  **e) All of the above**.

---

*(Class Discussion Notes)*
For the Hamiltonian $\hat{H} = C(\hat{S}_{1z} + \hat{S}_{2z})$, which representation is <u>convenient</u>? Uncoupled, coupled, both, or neither? Why?

For the Hamiltonian $\hat{H} = C(\hat{S}_{1z} + \hat{S}_{2z})$, the Hamiltonian is diagonal in both the **coupled and uncoupled representations**. Both the **coupled and uncoupled representation** are <u>convenient</u>.

---